\documentclass[%
 reprint,
 amsmath,amssymb,
 aps,
]{revtex4-2}

\usepackage{amsmath}

\usepackage{graphicx}% Include figure files
\usepackage{dcolumn}% Align table columns on decimal point
\usepackage{bm}% bold math

\begin{document}
\preprint{APS/123-QED}

\title{Scattering Singularity in 
Topological Dielectric Photonic Crystals}% Force line breaks with \\

\author{Langlang Xiong$^{1,4}$}

\author{Xunya Jiang$^{1,2,3}$}%
\email{jiangxunya@fudan.edu.cn}

\author{Guangwei Hu$^4$}
 \email{guangwei.hu@ntu.edu.sg}

 \affiliation{$^1$
Institute of Future Lighting, Academy for Engineering and Technology, Fudan University, Shanghai 200433, China}%Lines break automatically or can be forced with \\

 \affiliation{$^2$Department of Illuminating Engineering and Light Sources, School of Information Science and Engineering, Fudan University,
Shanghai 200433, China}%Lines break automatically or can be forced with \\

\affiliation{$^3$Engineering Research Center of Advanced Lighting Technology, Fudan University, Ministry of Education, Shanghai 200433, China}

\affiliation{
 $^4$School of Electrical and Electronic Engineering, 50 Nanyang Avenue, Nanyang Technological University, Singapore, 639798, Singapore
}%

\date{\today}% It is always \today, today,

\begin{abstract}

The exploration of topology in natural materials and metamaterials has garnered significant attention. Notably, the one-dimensional (1D) and two-dimensional (2D) Su-Schrieffer-Heeger (SSH) model, assessed through tight-binding approximations, has been extensively investigated in both quantum and classical systems, encompassing general and higher-order topology. Despite these advancements, a comprehensive examination of these models from the perspective of wave physics, particularly the scattering view, remains underexplored. In this study, we systematically unveil the origin of the 1D and 2D Zak phases stemming from the zero-scattering point, termed the scattering singularity in k-space. Employing an expanded plane wave expansion, we accurately compute the reflective spectrum of an infinite 2D photonic crystal (2D-PhC). Analyzing the reflective spectrum reveals the presence of a zero-scattering line in the 2D-PhC, considered the topological origin of the non-trivial Zak phase. Two distinct models, representing omnidirectional non-trivial cases and directional non-trivial cases, are employed to substantiate these findings. Our work introduces a novel perspective for characterizing the nature of non-trivial topological phases. The identification of the zero-scattering line not only enhances our understanding of the underlying physics but also provides valuable insights for the design of innovative devices.
\begin{description}
\item[Keywords]
Topological Singularity, Scattering Analysis, Extended Plane Wave Expansion, Photonic Crystals, Topological Phase Transition
\end{description}
\end{abstract}

\maketitle

%\tableofcontents

\section{Introduction}

The exploration of topological phenomena in quantum and classical systems \cite{qi2011topological,hasan2010colloquium, bansil2016colloquium,chiu2016classification,ozawa2019topological,rider2019perspective,lu2014topological,khanikaev2017two,wu2017applications,hu2020topological,xiong2021resonance} has garnered widespread interest due to their ability to host robust states, exemplified by edge states \cite{kim2020recent,xie2021higher,liu2021bulk,xie2018second,xie2019visualization,xiong2022topological} and corner states \cite{benalcazar2017quantized,he2020quadrupole,benalcazar2017electric}.  In the realm of topology, a classic analogy depicts a donut or a coffee mug, both possessing one hole, as homeomorphic, categorizing this as non-triviality. In contrast, a holeless cow and a sphere, both lacking holes, are deemed trivial \cite{Topology}. The band topology of an isolated band in quantum and classical systems can be likened to finding a ``hole" in a band. For a non-trivial band, discernible topological invariants such as the Zak phase \cite{zak1989berry, liu2017novel} or Chern number \cite{zhao2020first, xi2020topological, wang2008reflection} can be identified. Most topological systems explored thus far have originated from tight-binding models, notably the one-dimensional (1D) and two-dimensional (2D) Su-Schrieffer-Heeger (SSH) model \cite{asboth2016short, xie2018second, liu2017novel}. Whether in the context of 1D or 2D SSH models, the manipulation of topological phases is readily achievable by adjusting intra-cell and inter-cell coupling strengths. This simplicity, facilitated by a straightforward Hamiltonian, enables the comprehensive calculation of topological properties. 

However, classical wave systems, such as photonic systems, present a unique challenge due to non-negligible long-range coupling dictated by Maxwell's equations. This intricate coupling cannot be succinctly described by a simple tight-binding Hamiltonian \cite{li2019singularity, joannopoulos2011photonic}. While analogous phenomena may be observed in both wave and tight-binding systems, elucidating a distinctive topological origin in wave systems poses a significant hurdle. Specifically, the question arises: can we accurately pinpoint the ``hole" in a band using the classical wave method, specifically through the lens of scattering?

\begin{figure*}[htbp]
	\centering
		\includegraphics[width=0.8\textwidth]{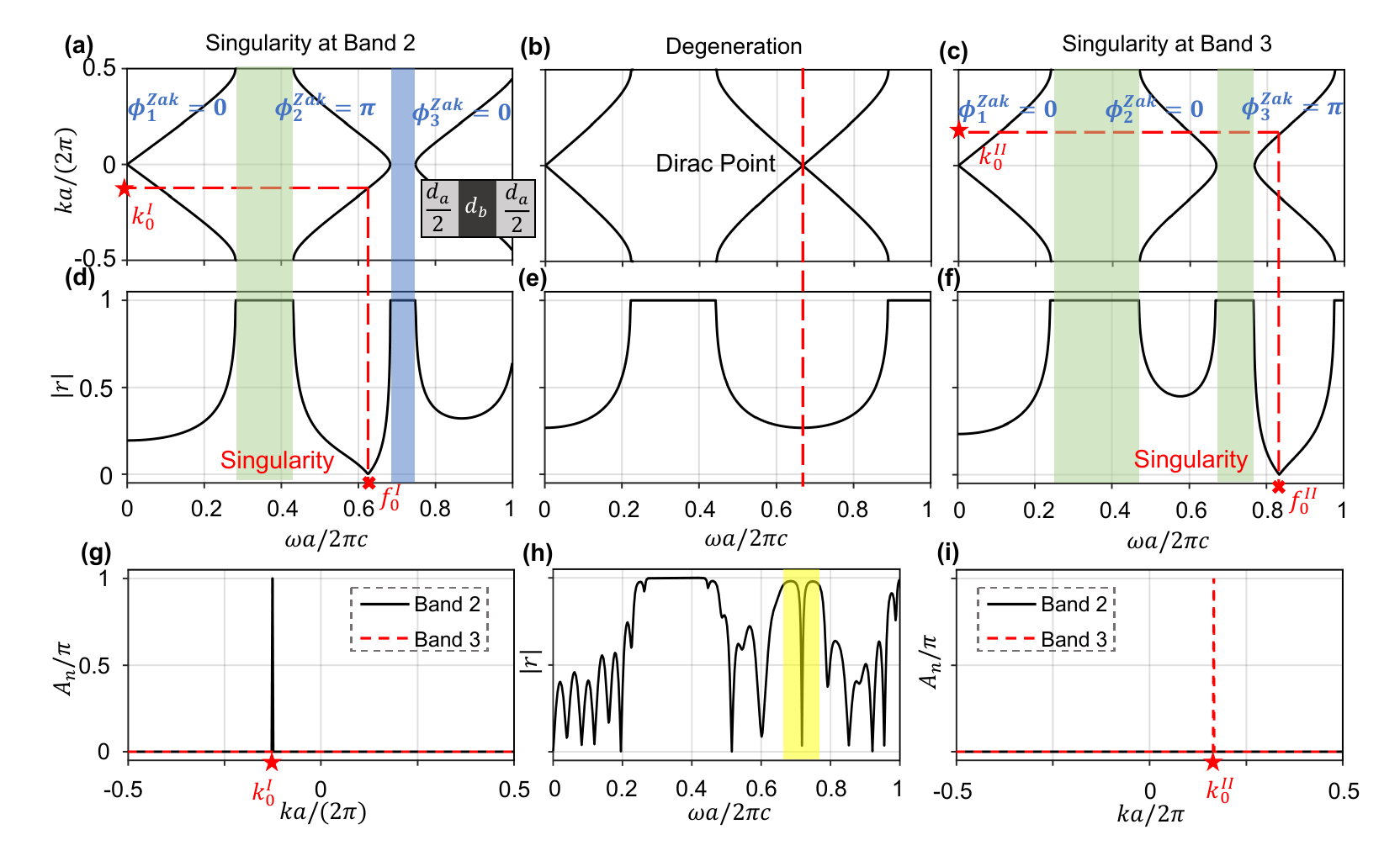}
	\caption{
 Singularity-induced topological phase transition in 1D-PhC. (a), (d), and (g): Band structure, reflective spectrum, and $A_n$ distribution, respectively, illustrating the singularity at band 2 with parameters $d_a=0.6a, d_b=0.4a, \varepsilon_a=1, \varepsilon_b=4$. The blue range represents the perfect magnetic conductor (PMC) gap, and the green range indicates the perfect electrical conductor (PEC) gap; (b) and (e): Band structure and reflective spectrum in the case of band degeneracy with parameters $d_a=0.75a, d_b=0.25a, \varepsilon_a=1, \varepsilon_b=9$; (c), (f), and (i): same as (a), (d), and (g), respectively, but showcasing the singularity at band 3 with parameters $d_a=0.8a, d_b=0.2a, \varepsilon_a=1, \varepsilon_b=9$; (h): Reflective spectrum of the structure formed by 3 layers of the model in (a) and 5 layers of the model in (c), with the edge state marked in the yellow range.}
 \label{fig1}
\end{figure*}

In this study, we first employ the transfer matrix method \cite{xiao2014surface} to precisely correlate the topological scattering singularity in 1D photonic crystals (PhCs)—a zero-reflection point within the reflective spectrum—with the Zak phase. Subsequently, our focus shifts to 2D-PhCs. To address the challenge of calculating the scattering of a half-infinite 2D-PhC, we meticulously introduce the expanded plane wave expansion method (EPWE) \cite{hsue2005extended, hsue2004applying}. Our analysis delves into two scenarios in 2D-PhCs. First, the two-rods square models without $C_4$ and $M_{x(y)}$ symmetries serve to illustrate the singularity-induced gap-closed-reopened process and topological phase transition \cite{xiong2022topological}. Second, the four-rods square models with $M_{x(y)}$ symmetry showcase the singularity behavior in directional Zak phase \cite{xie2018second}. This comprehensive approach enhances our understanding of the topological scattering singularity in 2D-PhCs. Our work introduces a novel perspective for considering the topological origin in classic systems, holding promise for advancing the development of new topological photonic devices.

\section{Scattering Singularity in 1D-PhC}

In this section, we will systematically define the concept of scattering singularity and elucidate its impact on band topology in 1D-PhC. While the most mathematical derivations are provided in the appendix of this work\cite{xiao2014surface}, this section aims to focus on physical insights.

Under the Transfer Matrix Method (TMM) gauge in a 1D-PhC, it can be proven that if an isolated band (excluding the 1st band) encompasses the resonance frequency point of the unit cell, then the Zak phase of this band must be $\pi$, designating such a zero-reflective resonance frequency point as a singularity. Consider, for example, the simplest model of a 1D-PhC with lattice constant $a$, as illustrated in Fig. \ref{fig1}. In this model, the unit cell comprises two types of layers: layer A with relative permittivity $\varepsilon_a$, permeability $\mu_a$, and thickness $d_a$, and layer B with relative permittivity $\varepsilon_b$, permeability $\mu_b$, and thickness $d_b$. Such a unit cell is also known as an Fabry–Pérot (FP) cavity, with a resonance condition defined as $\sin(k_b d_b) = 0$. Consequently, the resonance frequency is given by $f = m \cdot c / (2n_bd_b)$, where $m = 1, 2, 3...$, $k_b = n_b\omega/c$ is the wave vector, $\omega = 2\pi f$ is the angular frequency, and $n_a = \sqrt{\varepsilon_a \mu_a}$ represents the refractive index of layer A.

To visually observe how the singularity alters the band topology, we consider three parameter sets where the singularity moves from the lower band to the degenerate point and eventually to the upper band. In Case I (Fig. \ref{fig1}(a) and (d)), the singularity resides in the lower 2nd band with parameters $d_a=0.6a, d_b=0.4a, \varepsilon_a=1, \varepsilon_b=4$. In Case II (Fig. \ref{fig1}(b) and (e)), the singularity is at the degenerate Dirac point between the 2nd and 3rd bands with parameters $d_a=0.75a, d_b=0.25a, \varepsilon_a=1, \varepsilon_b=9$. Finally, in Case III (Fig. \ref{fig1}(c) and (f)), the singularity is at the 3rd band with parameters $d_a=0.8a, d_b=0.2a, \varepsilon_a=1, \varepsilon_b=9$. Clearly, as the singularity moves, the second gap undergoes closure and reopening, a hallmark of topological phase transition. Notably, when the singularity is at the second band, the Zak phase of the second band is $\pi$, while that of the third band is $0$. Conversely, when the singularity is at the third band, the results are reversed. The nature of the second gap also changes, transitioning from a non-trivial perfect magnetic Conductor (PMC)-like gap to a trivial perfect electrical conductor (PEC)-like gap. Consequently, combining the two photonic crystals reveals the presence of an edge state in the second gap, as illustrated in Fig. \ref{fig1}(h). This prompts the question: how the scattering singularity and band topology are connected? To answer this, we must revert to the definition of Zak phase for an isolated band:

\begin{equation}
    \phi_n^{Zak} = \int_{\rm{FBZ}} d k\ i \left\langle u_{n, k} \mid \partial_k u_{n, k}\right\rangle,
\end{equation}
where $\mid u_{n, k_j}\rangle$ is the normalized periodic part for $n$-th band at $k_j$ (Bloch wave-vector) of the field in a cell, and FBZ means the first Brillouin zone. The Zak phase can be numerically calculated using the Wilson loop \cite{normalized}:

\begin{equation}
\begin{aligned}
    \phi_n^{Zak} &= \sum_{j \in {\rm{FBZ}}}  A_n \\ &= \sum_{j \in {\rm{FBZ}}} -\textbf{Im} \{\ln \left[\left\langle u_{n, k_j} \mid u_{n, k_{j+1}}\right\rangle\right]\},
    \end{aligned}
\end{equation}
it's noteworthy that the product $A_n$ does not depend on the phase of $\mid u_{n, k_j}\rangle$s. In Fig. \ref{fig1}(g) and (i), we display each $A_n$ for the 2nd band and 3rd band in FBZ for case I and case III, respectively. Notably, at the point $(f_0^{I},k_0^{I})$ of case I, there is a sudden change of $A_n$, and similarly, at the point $(f_0^{III},k_0^{III})$ of case III, a sudden change is observed in the $A_n$. This phenomenon can be elucidated by the fact that $\mid u_{n, k_0^{+}}\rangle = -\mid u_{n, k_0^{+}}\rangle$, providing an explanation for the observed phase change at the singularity \cite{xiao2014surface}. Furthermore, within the context of TMM gauge, it is essential to emphasize that the group velocity at the singularity must be positive, ensuring the existence of only one singularity at a fixed frequency.

\section{Scattering Singularity in 2D-PhC}

Unlike the perfect gauge of TMM in 1D-PhC, when transitioning to 2D systems, certain challenges arise in researching the singularity. Firstly, the challenge of finding points that fulfill $\mid u_{n, k_{0}^+}\rangle = -\mid u_{n, k_{0}^-}\rangle$ since a random phase will be introduced when calculate the eigen-function of 2D-PhC. Fortunately, this can be addressed in the calculation of the Zak phase, thanks to the gauge independence of the Wilson loop, leading to the cancellation of the random phase. What's more, locating the singularity from a scattering perspective proves challenging, as computing the scattering of an $half$-$infinite$ 2D-PhC is a complex task. In the following section, we will introduce the EPWE to overcome computational challenges associated with the scattering of a half-infinite 2D-PhC and elucidate the connection between scattering singularities and directional Zak phase in 2D-PhC.

\subsection{Method: Expanded Plane Wave Expansion Method}

\begin{figure*}[htbp]
	\centering
		\includegraphics[width=0.9\textwidth]{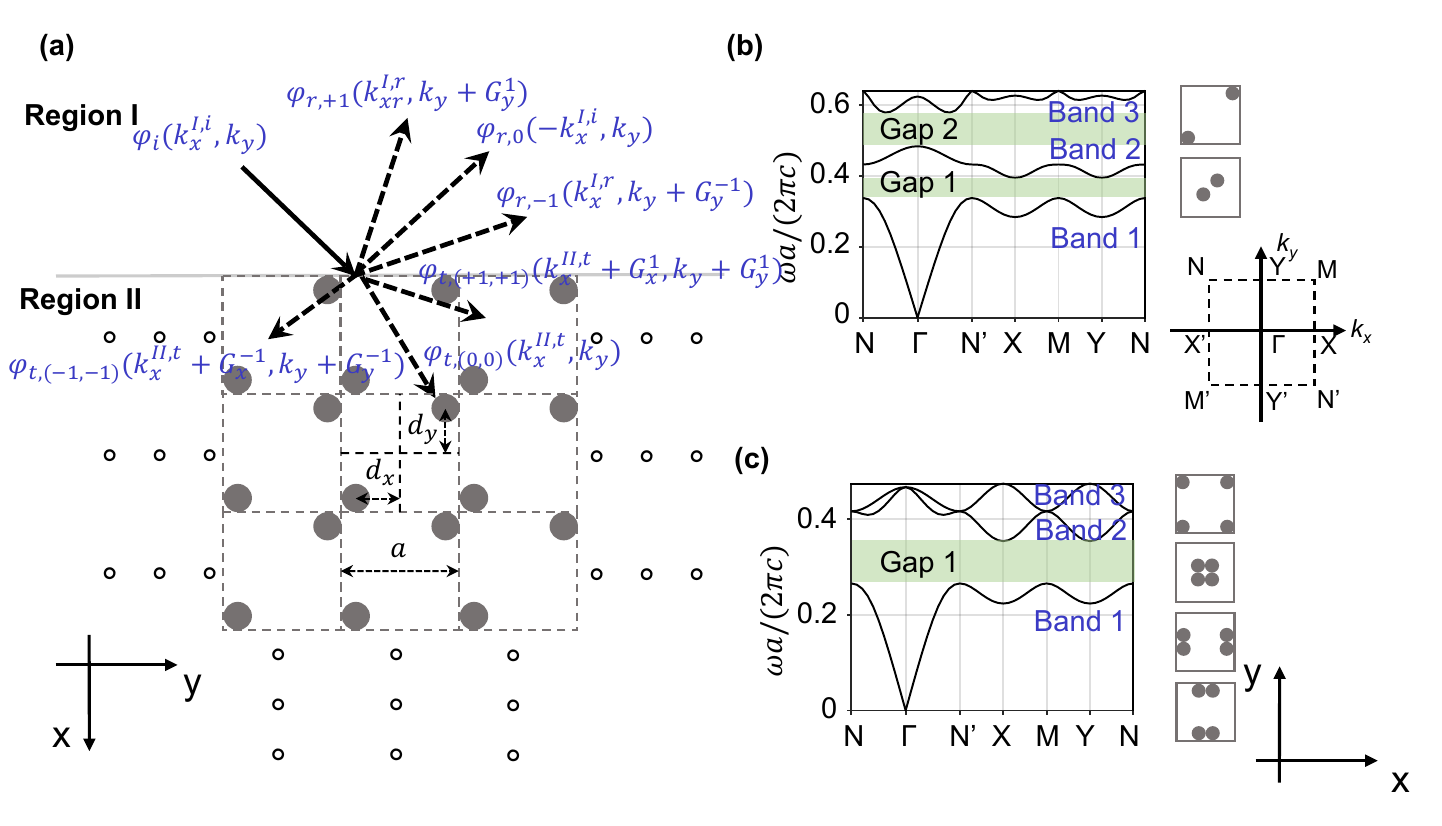}
	\caption{
 (a) Schematic representation of the structure and EPWE method; (b) Band structure of the two-rods model, along with the first Brillouin zone, where the two units cells on the right exhibit the same band structure in the left; (c) Band structure of the four-rods model, where the four unit cells in the right share the same band structure in the left;
 }
\label{fig2}
\end{figure*}

Considering a non-magnetic medium and taking Transverse-Magnetic (TM) polarization ($E^z$) as an example, the governing equation for the master function of Maxwell's equations is expressed as:

\begin{equation}
    \frac{1}{\varepsilon(\mathbf{r})} \nabla \times \nabla \times E^z(\mathbf{r}, \omega)=k_0^2 E^z(\mathbf{r}, \omega),
    \label{masfun}
\end{equation}
where $k_0 = \omega/c$. We expand $E^z$ and $\varepsilon(\mathbf{r})$ in Fourier space as $E^z(\mathbf{r})=\sum_{\mathbf{G}} e^{i(\mathbf{k}+\mathbf{G}) \cdot \mathbf{r}} E^z_{\mathbf{k}, \mathbf{G}}$ and $\varepsilon(\mathbf{r})= \sum_{\mathbf{G}} e^{i\mathbf{G} \cdot \mathbf{r}} \varepsilon_{\mathbf{G}}$, where $\mathbf{G}$ is the reciprocal lattice vector. The Equ. (\ref{masfun}) is then rewritten in Fourier space as:
\begin{equation}
    \sum_{\mathbf{G}^{\prime}} \varepsilon_{\mathbf{G}-\mathbf{G}^{\prime}}^{-1}(\mathbf{k}+\mathbf{G}) \cdot\left(\mathbf{k}+\mathbf{G}^{\prime}\right) E^z_{\mathbf{k}, \mathbf{G}^{\prime}} = k_0^2 \sum_{\mathbf{G}^{\prime}} E^z_{\mathbf{k}, \mathbf{G}^{\prime}},
    \label{PWE}
\end{equation}
Equ. (\ref{PWE}) represents an eigen-function, allowing the determination of the band structure and eigen-field of a 2D-PhC. 
However, it remains insufficient for solving the scattering problem due to challenges in defining an ``interface" and an input source. To address this, we rewrite Equ. (\ref{PWE}) as:

\begin{widetext}
    \begin{equation}
        \left[\begin{array}{cc}
        0 & \hat{\mathbf{I}} \\
        \boldsymbol{\varepsilon}_{\mathbf{G}-\mathbf{G}^{\prime \prime}}
        \left[
        k_0^2 -\boldsymbol{\varepsilon}_{\mathbf{G}^{\prime \prime}-\mathbf{G}^{\prime}}^{-1}\left(\mathbf{G}^{\prime \prime}+k_y \hat{\mathbf{y}}\right) \cdot\left(\mathbf{G}^{\prime}+k_y \hat{\mathbf{y}}\right)
        \right] 
        &  -\boldsymbol{\varepsilon}_{\mathbf{G}-\mathbf{G}^{\prime \prime}} \left[ \boldsymbol{\varepsilon}^{-1}_{\mathbf{G}^{\prime \prime}-\mathbf{G}^{\prime}}
        
        \left(G_x^{\prime \prime}+G_x^{\prime} \right) 
        \right] 
        \end{array}\right] 
        \left[\begin{array}{c} E_{\mathbf{G}^{\prime}} \\
        k_x E_{\mathbf{G}^{\prime}} \end{array}\right]
        =k_x\left[\begin{array}{l} E_{\mathbf{G}^{\prime}} \\ k_x E_{\mathbf{G}^{\prime}}
        \end{array}\right],
        \label{EPWE}
    \end{equation}
\end{widetext}
where $\mathbf{k}=k_x \hat{\mathbf{x}} + k_y \hat{\mathbf{y}}$. In this equation, $k_y$ and $k_0$ is fixed, and the determined parameter is $k_x$. This equation is suitable when the interface lies along the y-direction (see Fig. \ref{fig2}(a)). A similar equation can be derived for scenarios where the interface is along the x-direction, in which case $k_x$ and $k_0$ are fixed, and the determined parameter is $k_y$. This equation is suitable for addressing inclined incidences from a homogeneous medium (region I) to a half-infinite 2D-PhC (region II), as illustrated in Fig. \ref{fig2}(a). By applying boundary condition, the relationship between the $E^z$ fields in region I and region II can be established as follows:
\begin{widetext}
\begin{equation}
\left[\begin{array}{cc}
-\left\langle x_0 y \mid E_m^{z,\mathrm{I}}\right\rangle & \left\langle x_0 y \mid E_m^{z,\mathrm{II}}\right\rangle \\
-\left\langle x_0 y\left|\partial_x\right| E_m^{z,\mathrm{I}}\right\rangle & \left\langle x_0 y\left|\partial_x\right| E_m^{z,\mathrm{II}}\right\rangle
\end{array}\right]\left[\begin{array}{c}
\left\langle E_m^{z,\mathrm{I}}|\hat{\mathbf{r}}| E_0^{z,\mathrm{I}}\right\rangle \\
\left\langle E_m^{z,\mathrm{II}}|\hat{\mathbf{t}}| E_0^{z,\mathrm{I}}\right\rangle
\end{array}\right] 
=\left[\begin{array}{c}
\left\langle x_0 y \mid E_0^{\mathrm{I}}\right\rangle \\
\left\langle x_0 y\left|\partial_x\right| E_0^{\mathrm{I}}\right\rangle
\end{array}\right],
\label{RT}
\end{equation}
\end{widetext}
where $\hat{\mathbf{r}}$ and $\hat{\mathbf{t}}$ are the reflection and transmission operators. $E_m^{I(II)}$ represents the m-th reflected (transmitted) mode in region I (II), and $E_0^{I}$ is the incident field.

In this equation, $E_m^{II}$ can be solved by Equ. (\ref{EPWE}). Regarding the eigenvectors in region I with homogeneous media, they are represented by plane waves and solely depend on $G_y$. Specifically, assuming the input wave-vector is $\mathbf{k}_i=[k_x^{I,i},k_y]$, according to Bloch theory, the reflection wave-vector is $\mathbf{k}_r=[-\sqrt{k_0^2-(k_y+G_y)^2}, k_y+G_y]$, and the transmission wave-vector is $\mathbf{k}_{t}=(k_x^{II,t}+G_x,k_y+G_y)$, where $k_x^{II,t}$ also can be solved by using Equ. (\ref{EPWE}). Therefore, if we have knowledge of the structure function $\varepsilon(\mathbf{r})$, the incident wave-vector $\mathbf{k}_i$ and the frequency of incident plane-wave $\omega$, we can solve the interface problem between a half-infinite homogeneous media and a half-infinite 2D-PhC by utilizing Equ. (\ref{EPWE}) and Equ. (\ref{RT}). 
We should note that, the reflective spectrum, calculated using EPWE, is projected in a specific direction. For instance, when the interface is perpendicular to the x-direction, as shown in Fig. \ref{fig2}(a), we exclusively consider specific modes in which the parallel wave vector component $k_y$ matches that of the incident light wave. This requirement stems from the conservation of $k_y$ on both sides of the interface. As a result, the reflective spectrum can be denoted as $R_x(k_y)$ or $R_y(k_x)$, signifying the reflectivity from the interface is perpendicular to the x-direction or y-direction, respectively.

\begin{figure*}[htbp]
	\centering
		\includegraphics[width=0.8\textwidth]{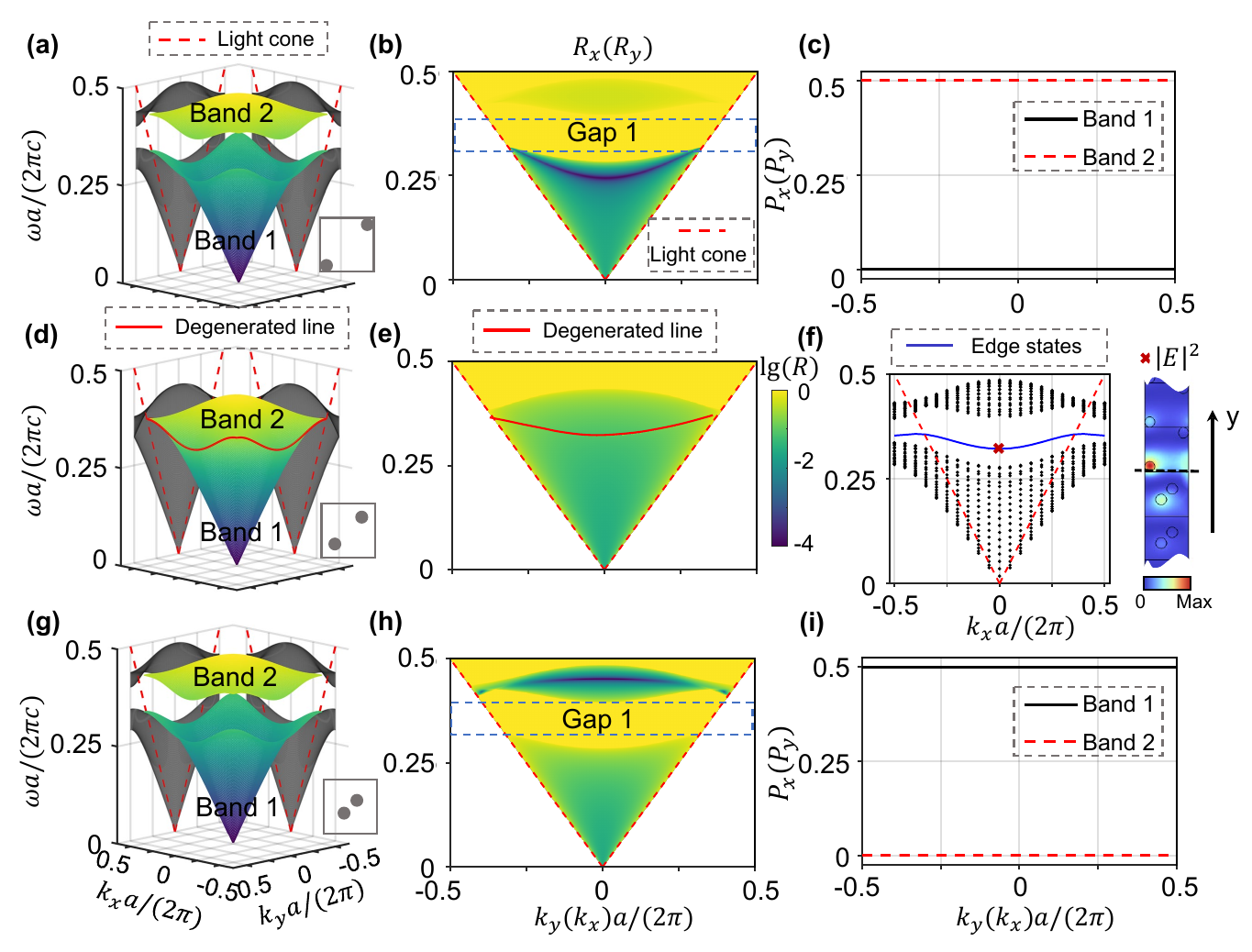}
	\caption{
 Singularity-induced topological phase transition in 2D-PhC. (a-c) Band structure in three dimensional and the projected band structure in x(y) direction, reflective spectrum $R_x(k_y)$ or $R_y(k_x)$, and the directional polarization of the two lowest bands, respectively, where the parameters are $r=0.12a, \varepsilon=12, d_x=d_y=0.12a$; (d-e) The same as (a-b) but $d_x=d_y=0.25a$; (g-i): The same as (a-c) but $d_x=d_y=0.38a$; (f): The band structure that combines 10 layers of the model in (a) with 10 layers of the model in (g), along with the $|E|^2$ distribution, where the band of edge states is marked with blue line.
 }\label{fig3}
\end{figure*}

\subsection{Results}

In this sub-section, we will investigate two cases to illustrate topological singularities in 2D-PhC using EPWE. In the first case, depicted in Fig. \ref{fig2}(b), the two-rods model lacking $C_4$ and $M_{x(y)}$ symmetries exhibits two clearly defined lowest bands without degeneracy. This characteristic makes it an ideal model for investigating the topological phase transition between these two lowest bands. In the second case, our focus is on examining the topological phase of the first band in different directions within a four-rods model, as shown in Fig. \ref{fig2}(c). In both cases, the rods with a radius of $r=0.12a$ are composed of $\varepsilon=12$ material and are situated in the air. Additionally, TM polarization is considered. It is worth noting that the right two unit cells in Fig. \ref{fig2}(b) and the right four unit cells in Fig. \ref{fig2}(c) have the same band structures, respectively, because only the centering point of the unit cells is changing. For an infinite bulk crystal where the interface is neglected, their eigen-frequencies are the same.

Before presenting the results, we should highlight the calculation of directional Zak phase in 2D-PhC \cite{xie2018second, liu2017novel}. While akin to the computation of 1D Zak phase in section II, it involves the integration of the Berry connection in two directions

\begin{equation}
    \phi_{l,n}^{Zak} = \int_{\rm{FBZ}} d k_x d k_y \ i\left\langle u_{n, k} \mid \partial_{k_l} u_{n, k}\right\rangle, l=x,y.
\end{equation}
We can alternatively utilize the Wilson loop to obtain the 2D Zak phase. For instance, selecting a fixed $k_y$, and the Zak phase in the x-direction at this fixed $k_y$ is given by:

\begin{equation}
    \phi_{x,n}^{Zak}(k_y) = \sum_{j \in {\rm{FBZ}}} -\textbf{Im}  \ln \left[\left\langle u_{n, k_{x,j}} \mid u_{n, k_{x,j+1}}\right\rangle\right].
\end{equation}
Additionally, directional polarization can be employed to ascertain topology, denoted as, $\mathbf{P_n}=[P_{x,n},P_{y,n}]=[\phi_{x,n}^{Zak}, \phi_{y,n}^{Zak}]/(2\pi)$. 
Consequently, we can correlate the directional polarization $P_x(k_y)$ or $P_y(k_x)$ to the reflective spectrum $R_x(k_y)$ or $R_y(k_x)$, respectively.

\begin{figure*}[htbp]
	\centering
		\includegraphics[width=0.8\textwidth]{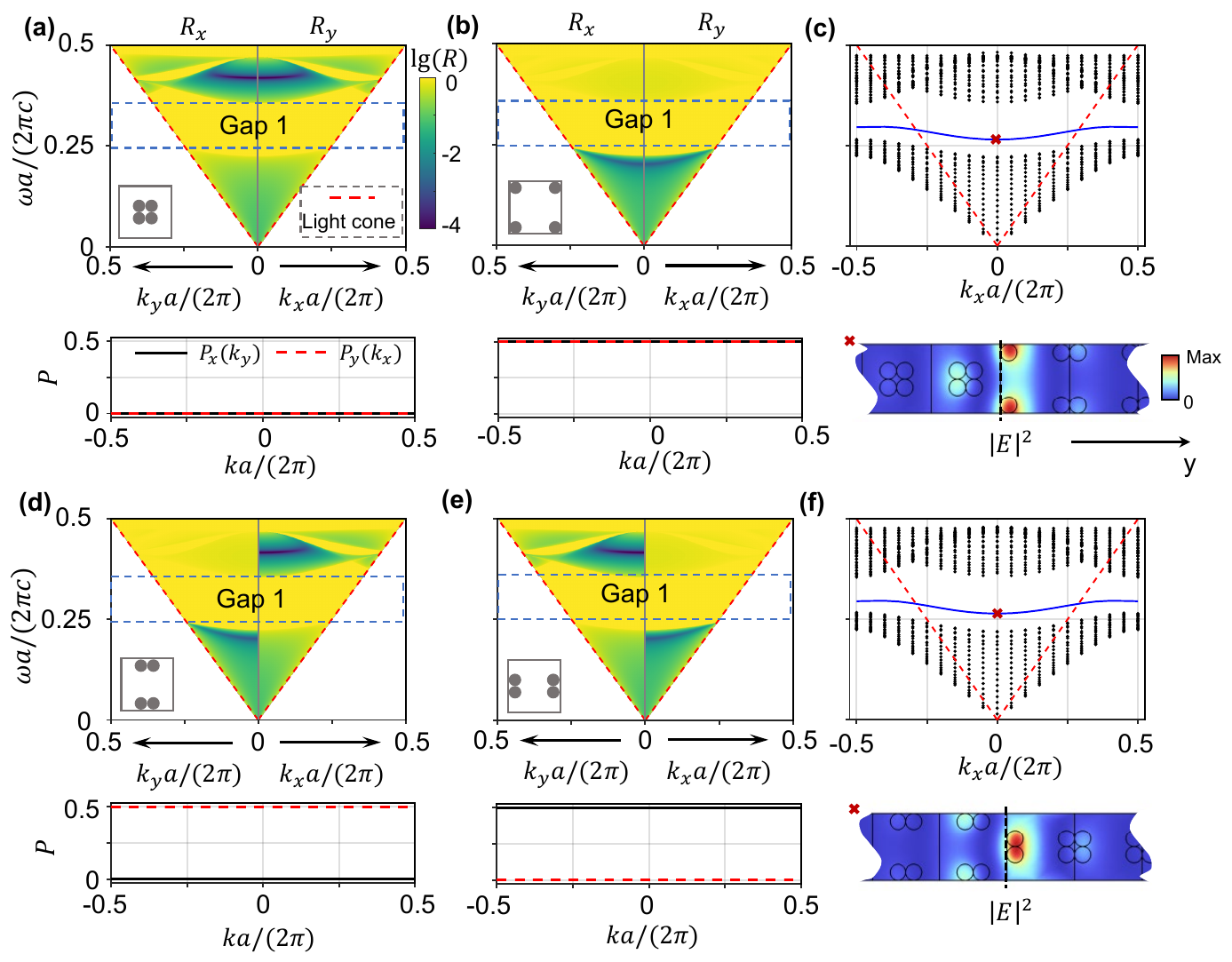}
	\caption{
 Singularity Behavior in Directional Zak Phase. (a) Top: reflective spectrum $R_x(k_y)$ along $k_ya/(2\pi)$ from 0 to 0.5 (left) and $R_y(k_x)$ along $k_xa/(2\pi)$ from 0 to 0.5 (right). Bottom: directional polarization of the first band in different directions, with parameters $d_x=d_y=0.12a$; (b) the same as (a) but $d_x=d_y=0.38a$; (c) Band structure combining 10 layers of the model in (a) with 10 layers of the model in (b), along with the $|E|^2$ distribution. (d) the same as (a) but $d_x=0.12a, d_y=0.38a$; (e) the same as (a) but $d_x=0.38a, d_y=0.12a$; (f) Band structure combining 10 layers of the model in (d) with 10 layers of the model in (e), along with the $|E|^2$ distribution.
}\label{fig4}
\end{figure*}

\subsubsection{Singularity-Induced Topological Phase Transition Between Two Bands}

Similar to the topological phase transition discussed in Section II, we construct a square lattice to break $C_4$ and $M_{x(y)}$ symmetries, then by varying the distance between the rods and the cell center, we investigate the entire topological phase transition in the lowest two bands. Maintaining equal distances between the two rods and the cell center $d_x=d_y$, so the band structure and band topology remain identical in both x and y directions. Therefore, the x-label in Figs. \ref{fig3}(b, e, h, c, i) can represent either $k_x$ or $k_y$. Accordingly, the reflective spectrum in Figs. \ref{fig3}(b, e, h) can indicate the corresponding $R_y(k_x)$ or $R_x(k_y)$, respectively. Similarly, the polarization in Fig. \ref{fig3}(c, i) can represent either $P_y(k_x)$ or $P_x(k_y)$, respectively.

In the case presented in Figs. \ref{fig3}(a-c), where the rods are positioned far away from the cell center ($d_x=d_y=0.38a$), the band topology in Fig. \ref{fig3}(c) reveals that $\mathbf{P_1}=[0.5,0.5]$ and $\mathbf{P_2}=[0,0]$. The topological scattering singularity in this scenario is particularly intriguing, with the zero-reflection phenomenon no longer manifesting as a point but as a line in the first band, as shown in Fig. \ref{fig3}(b) with deep-blue line. This observation aligns with the directional polarization in Fig. \ref{fig3}(c): for any fixed  $k_x$ or $k_y$, if $P_{1,x}(k_y)=0.5$ or $P_{1,y}(k_x)=0.5$, the projected reflective spectrum $R_x(k_y,\omega)$ or $R_y(k_x,\omega)$ exhibits a zero-reflection line.

By varying $d_x$ and $d_y$, we can induce a topological phase transition through a gap-close-reopen process. In Figs. \ref{fig3}(d-e), the first gap closes when $d_x=d_y=0.25a$. Further decreasing $d_x$ and $d_y$ to $0.12a$ reopens the first gap, as shown in Fig. \ref{fig3}(g), and the topology of the lowest bands becomes opposite to the first case. In this scenario, there is no zero-reflection singularity in the first band of the projected reflective spectrum, and the zero-reflection singularity is observed in the second band. It's observed that the second band may overlap when projected to one direction, especially near the band edge, but this does not affect our conclusions.

Additionally, combining the two types of 2D-PhC in Fig. \ref{fig3}(a) and (g) to verify the topology, the band structure and field distribution of edge states are illustrated in Fig. \ref{fig3}(f). Notably, the band of edge states in the first gap is clearly discernible, along with the presence of a localized edge state with $k_x=0$.

\subsubsection{Singularity Behavior in Directional Zak Phase}
In this sub-section, we delve into the nuanced intricacies of the topological singularity's behavior within the directional Zak phase, offering a comprehensive exploration of its impact on the first band in a four-rods model. By systematically varying the parameters, such as the distances between rods ($d_x$ and $d_y$), we scrutinize the resulting changes in the reflective spectrum's symmetry and the emergence of zero-reflection lines. 

In cases where $d_x$ equals $d_y$, i.e., $d_x=d_y=0.12a$ or $d_x=d_y=0.38a$, we plot $R_y(k_x)$ and $R_x(k_y)$ that along different directions, as portrayed in Figs. \ref{fig4}(a-b). The reflective spectrum exhibits notable symmetrical characteristics along $k_x(k_y)=0$, culminating in a reflective symmetry axis. Our analysis discerns the absence of a zero-reflection line in the trivial first band, contrasting sharply with the discernible ze ro-reflection line in the non-trivial first band. The nuanced symmetry deviations are further underscored by the pronounced presence of edge states in the combined models of these two distinct photonic crystals.

The exploration extends to scenarios where $d_x\ne d_y$, i.e., $d_x=0.12a, d_y=0.38a$ or $d_x=0.38a, d_y=0.12a$, as exemplified in Figs. \ref{fig4}(d-e). In such instances, the once symmetrical reflective spectrum along $k_x(k_y)=0$ undergoes a transformative shift, eliminating the reflective symmetry axis. The non-trivial direction of the first band, where $P_x(P_y)\ne 0$, distinctly manifests a discernible zero-reflection line in $k_y(k_x)$ direction. In contrast, the trivial direction exhibits the absence of such a line. This observation is further validated by the unequivocal presence of edge states in the combined model of the two diverse photonic crystals.

From our two examples, it is important to note that even if two or more cells have the same band structure, their reflective spectra may differ fundamentally. Our work provides a scattering view to reveal the deep topology beyond the band structure.

\section{Conclusions}

In summary, this study has unveiled a novel approach for discerning the topology of isolated bands in classical wave systems, both in one and two dimensions, leveraging the concept of a topological singularity characterized by a zero-reflection point or line. Our investigations in the two-dimensional domain revealed that the dynamic movement of this topological singularity induces significant topological phase transitions. The existence of topological singularities in diverse directions imparts distinct topological properties along those specific axes. Our contributions extend beyond traditional tight-binding models, as demonstrated by the designed models showcasing topological phase transitions in two-dimensional photonic crystals. These findings underscore the profound impact of the topological singularity on reflective phenomena, shedding light on the intricate interplay between symmetry, zero-reflection lines, and band topology. The proposed methodology not only enriches our fundamental understanding of topological phases but also opens avenues for practical applications in the design and engineering of topological devices. As future work, we anticipate further exploration of topological aspects in diverse topological insulators, promising an extended trajectory of discoveries and applications in the burgeoning field of wave physics.

\section*{Data availability statement}
The code used to calculate the band and reflective spectrum is available at https://github.com/kabume/Peacock.jl

\begin{acknowledgments}
	This work is supported by National High Technology Research and Development Program of China (17-H863-04-ZT-001-035-01); National Key Research and Development Program of China (2016YFA0301103, 2018YFA0306201);  National Natural Science Foundation of China (12174073); G.H. acknowledges the Nanyang Assistant Professorship Start-up Grant and Ministry of Education (Singapore) under AcRF TIER1 (RG61/23).
\end{acknowledgments}

\section*{Disclosures}
The authors declare no conflicts of interest.

%\nocite{*}

\bibliography{apssamp}% Produces the bibliography via BibTeX.

\begin{thebibliography}{33}
\bibitem [{nor()}]{normalized}%
  \BibitemOpen
  \href@noop {} {\bibinfo {title} {It's essential to note that
  $\left[\left\langle u_{n, k_{j}} \mid u_{n, k_{j+1}}\right\rangle\right]$
  should be normalized with $\varepsilon(r)$ in all of our case.}}\BibitemShut
  {Stop}%
\end{thebibliography}

\end{document}